\documentclass[%
preprint,
 amsmath,amssymb,
 aps,
 prl,
 lengthcheck,%
]{revtex4-1}

\usepackage{graphicx}

\def\bfM{\mathbf{M}}
\def\bfe{\mathbf{e}}
\def\bfI{\mathbf{I}}

\begin{document}

\preprint{APS/123-QED}

\title{Continuum Electromechanical Modeling of Protein-Membrane Interaction}

\author{Y. C. Zhou}
\email{yzhou@math.colostate.edu}
\affiliation{Department of Mathematics, Colorado State University, Fort Collins, CO 80523, USA}%

\author{Benzhuo Lu}%
\affiliation{State Key Laboratory of Scientific/Engineering Computing, 
Institute of Computational Mathematics and Scientific/Engineering Computing, 
Chinese Academy of Sciences, Beijing 100190, China}%


\author{Alemayehu A. Gorfe}
\affiliation{Department of Integrative Biology and Pharmacology \\
University of Texas Medical School at Houston, Houston, TX 77225, USA}




\date{\today}

\begin{abstract}
A continuum electromechanical model is proposed to describe the membrane curvature induced by electrostatic
interactions in a solvated protein-membrane system. The model couples the macroscopic strain energy of membrane
and the electrostatic solvation energy of the system, and equilibrium membrane deformation is obtained by minimizing
the electro-elastic energy functional with respect to the dielectric interface. The model is illustrated with the
systems with increasing geometry complexity and captures the sensitivity of membrane curvature to the permanent and 
mobile charge distributions.
\end{abstract}

\maketitle
Boundaries of eukaryotic cells and most organelles within the cells are defined by lipid 
bilayer membranes. Shape and topological transformations of membrane are crucial steps 
in numerous transport and signaling processes of cells, including cell 
migration, membrane trafficking, and ion conductance \cite{RidleyA2003a,ChoW2005review,WigginsP2005a}. 
There are various types of proteins that are displaced to the vicinity of membranes or are embedded in them. 
Many of the delicate changes of configuration and topology of biological membranes result from the 
interactions between lipids as well as between lipids and proteins. Among the most interesting examples
of protein-induced membrane deformation are the Bin/amphiphysin/Rvs (BAR) domain-induced membrane 
curvature \cite{Peter04_science} and the Endosomal Sorting Complex Required for Transport III (ESCRT-III) 
induced membrane budding or protrusion \cite{SaksenaS2007a}. These interactions happen over a wide range of 
time and length scales, and depend critically on the lipid composition, charge distribution, 
hydrophobicity, and elastic moduli. Understanding of quantitative 
relationships between the forces and the membrane geometry have long been subjected to 
intensive studies using either discrete methods such as molecular dynamics (MD)
simulations and the coarsed grained MD \cite{BloodP2006}, or continuum elasticity based on
minimization of Helfrich-Canham-Evans bending free energy \cite{FygensonD1997a,FabrikantG2009a} and 
its simplifications \cite{ArkhipovA2008a}, sometimes under the constraints such as surface area or enclosed 
volume \cite{Klug2006a}. While the dynamics in the lipid-protein interactions can be revealed in atomistic
details by MD simulations, protein-induced macroscopic bending, expansion, contraction of lipid
bilayers can be more efficiently simulated using continuum mechanics 
at substantially reduced computational cost. The continuum models also give us access 
to interacting at the interfaces between molecular biology, mathematical modeling, 
and numerical computing. Most of the current studies of membrane mechanics model
the lipid bilayer as an elastic sheet with vanishing thickness, and are focused on the 
the membrane's curvature energy, stability, or the deformation modes for given 
external mechanical constraints or loads. In this Letter, we propose a continuum 
electromechanical energy and associated system of nonlinear partial differential equations 
as a framework for modeling the self-consistent macromolecular deformation induced by the 
long range electrostatic interactions in solvated macromolecular systems. 

\begin{figure}[!ht]
\begin{center}
\includegraphics[height=3.2cm]{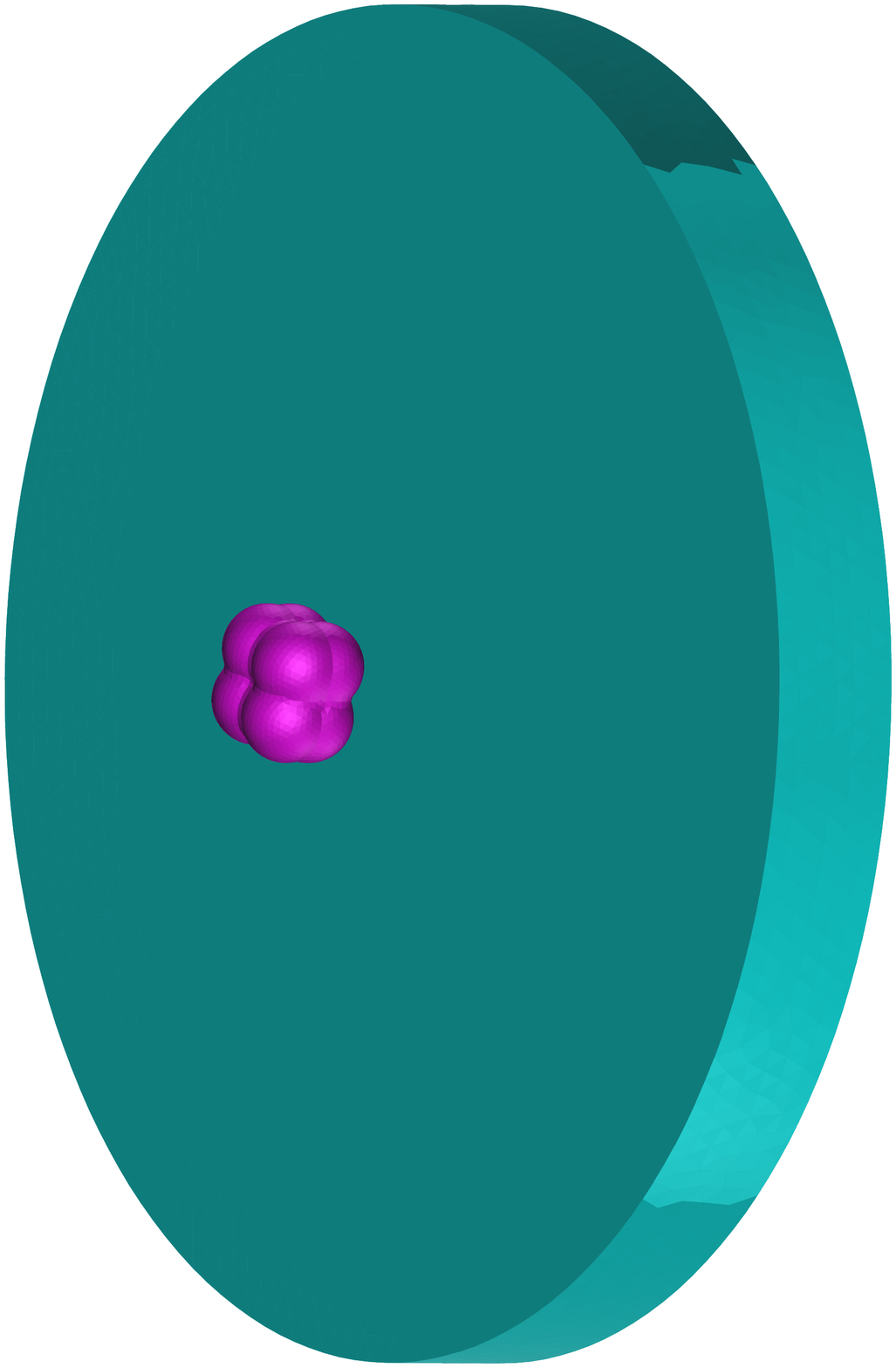} \hspace{5mm}
\includegraphics[height=3.2cm]{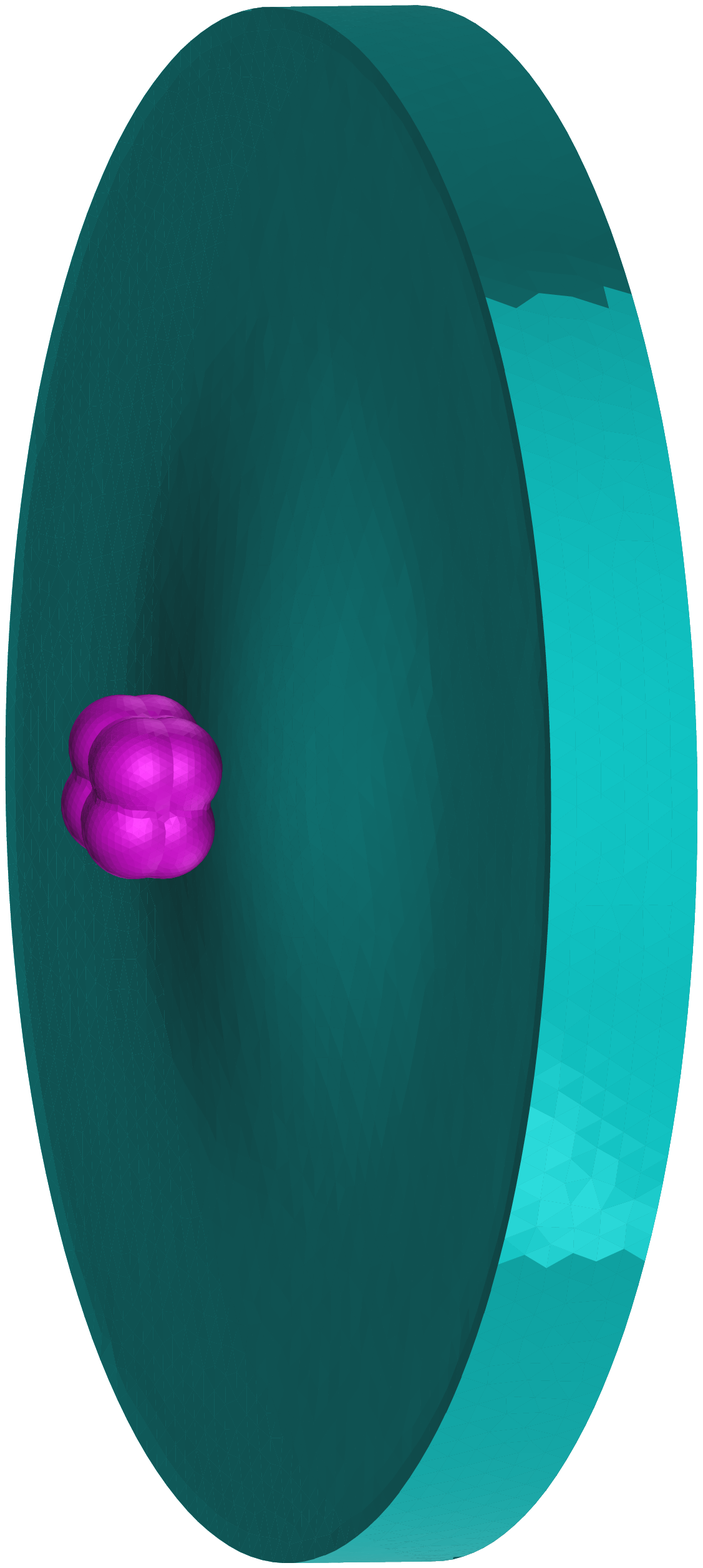} \hspace{5mm}
\includegraphics[height=3.2cm]{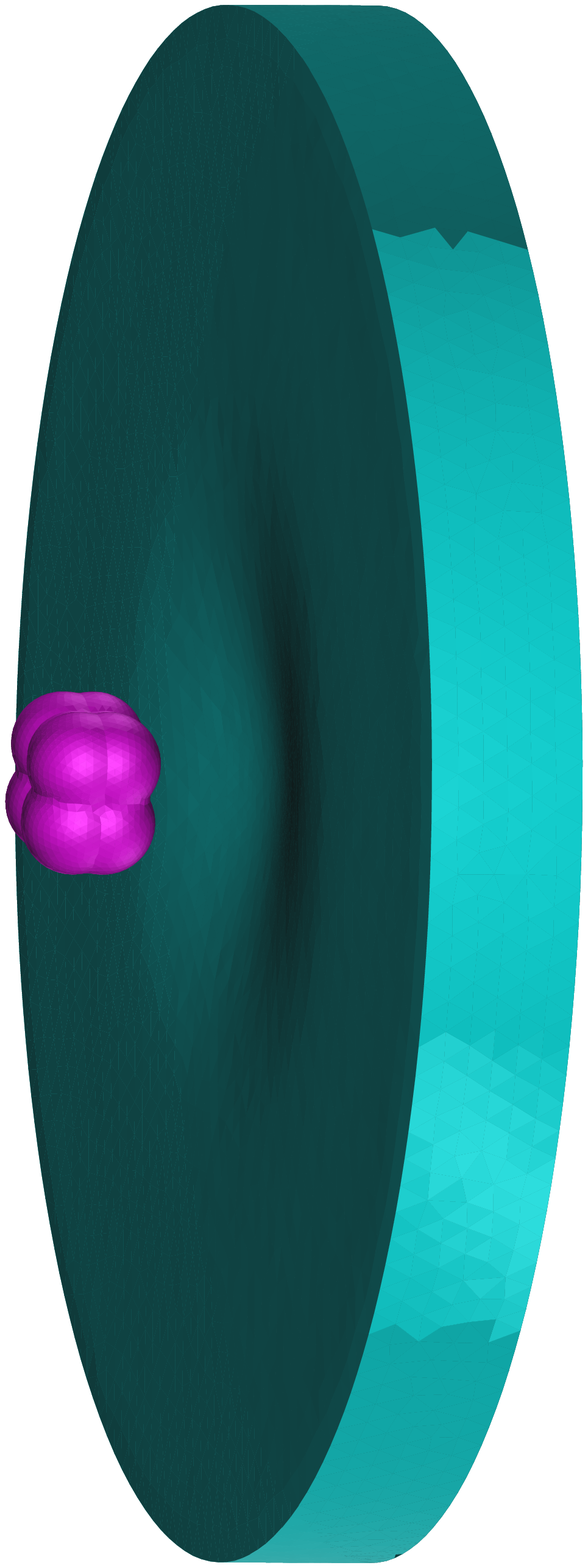}
\end{center}
\caption{Illustration of the proposed electromechanical model for protein-membrane interactions. Left: 
Molecular surface of a protein-membrane complex. Atomistic details are retained in the protein but 
are neglected in membrane. The left and right surfaces of the membrane are negatively charged, and its
circular face is charge free and constrained. Depending on the mobile ion concentration and the partial charge 
distribution in the protein, the membrane might be attracted to bend towards the protein (Middle) or repulsed 
away from the protein (Right).} 
\label{fig:illu_complex}
\end{figure}

The essential idea underlying our model is the coupling of the electrostatic potential
energy of the solvated macromolecular system and the nonlinear elastic energy of flexible
biomolecules modeled as elastic continuum. We stress that, as opposed to other continuum
models, the bilayer is described as an elastic sheet with a finite thickness and a constant dielectric permittivity. 
We neglect the atomistic details of the lipids, but the partial charges in the head-groups of the lipid bilayers, 
modeled as a distribution of charge on the surfaces, will be retained and incorporated into the modeling of the 
electrostatic interactions with the proteins, c.f. Fig.~\ref{fig:illu_complex}. In contrast, the proteins
are modeled as rigid bodies with atomistic detail. Their dielectric interfaces with the solvent are 
described by properly defined molecular surface.

Let $\phi$ be the electrostatic potential of the entire solvated system and $u$ be the displacement vector 
of its flexible subdomain with a mass density $\rho_m$, the electro-elastic energy density of the system is defined 
to be
\begin{equation} \label{eqn:total_energy}
\mathcal{F} = \frac{1}{2} \rho_m |\dot{u}|^2 + \frac{1}{2} \left [ \lambda (\sum_{i=1}^3 \bfe_{ii} ) ^2
+ 2 \mu \sum_{i,j=1}^3  \bfe_{ij}^2 \right] + g,
\end{equation}
where the first and the second terms are the kinetic and potential energy of the elastic
domain in the system, respectively, and $\bfe=\frac{1}{2} (\nabla u + \nabla u^T)$ is the linear 
strain tensor. $\lambda, \mu$ are the two Lam\'e constants. The electrostatic potential energy (density) is given by $g$,
for which we use the following ansatz: 
\begin{equation} \label{eqn:g_def}
g =-\frac{\epsilon}{2} |\nabla \phi|^2 - k_BT \sum_{i=1}^{N} c_i (e^{-q_i \phi/k_BT}-1) + \rho_q \phi, 
\end{equation}
where $\epsilon$ is the spatial-dependent dielectric permittivity, $c_i$ is the concentration of the $i$th species of 
mobile ions in the solvent with charge $q_i$. The permanent charge distribution $\rho_q$ consists of the $M$ singular 
point charges located at $x_i$ in the proteins and the 
constant charge distribution $\rho_s$ on the surfaces $S_m$ of the membrane, i.e.,
$\rho_q = \sum_{i=1}^M q_i \delta(x_i) + \rho_s [ \chi(S_m) ]$,
where $\delta$ and $\chi$ are the 3-D and 2-D Dirac delta functions, respectively. It is
known \cite{Sharp90} that this potential $\phi$ can be solved from the Poisson-Boltzmann (PB) equation,
$-\nabla \cdot (\epsilon \nabla \phi) - \sum_{i=1}^N c_i q_i e^{-q_i \phi/k_BT} = \rho_q$,
and the electrostatic force density $f=-\nabla g$ is given by
\begin{eqnarray} \label{eqn:eforce_def}
f & = & \rho_q E - \frac{1}{2} \left( \epsilon_s |E_s|^2 - \epsilon_m |E_m|^2\right) \chi (\Gamma) \nonumber \\
& & -k_B T \sum_{i=1}^N c_i ( e^{-q_i \phi/k_BT} -1) \chi (\Gamma),
\end{eqnarray}
where $E =-\nabla \phi$, $\Gamma$ is the molecular surface, and the subscripts $s,m$ denote the values in the
solvent and in the molecules, respectively. Here we choose $\epsilon_s = 80$ and $\epsilon_m = 1$. 
The surface charge distribution will be treated as an interface condition on the membrane surfaces, i.e.,
$-(\epsilon_s \nabla \phi_s - \epsilon_m \nabla \phi_m) \cdot n = \rho_s$,
where $n$ is the surface outer normal. We finally consider the energy functional 
\begin{equation} \label{eqn:Functional}
S = \int_{t_1}^{t_2} \int_{\Omega} \mathcal{F} dx dt 
\end{equation}
and its variation with respect to the spatial displacement $ \delta S = \int_{t_1}^{t_2} \int_{\Omega} \delta \mathcal{F} dx dt$.
The minimization of the energy functional gives rise to an equation
$\delta \mathcal{F} = 0$, i.e.,
\begin{equation} \label{eqn:elastic_eq}
 \rho_m \ddot{u} -
\nabla \cdot \big( \lambda \sum_{i=1}^3 \bfe_{ii} \bfI + 2 \mu \bfe \big) = f,
\end{equation}
where $\bfI$ is the identity matrix. This shows that the minimization of the electro-elastic energy 
functional (\ref{eqn:Functional}) is equivalent to the
coupled solution of the PB equation and the elastic equation (\ref{eqn:elastic_eq}). 
The coupling is realized through supplying the electrostatic force computed from the PB equation 
to the elastic equation as a dynamical load and supplying the varying membrane configuration computed
from elastic equation to the PB equation as a dielectric interface. The time derivative 
in the elastic equation can be removed if one is only interested in the equilibrium configuration of the deformable 
macromolecules. This regenerates the direct coupling of the nonlinear elasticity equation and the PB equation
\cite{ZhouY2007c}, and gives us a handle of high-fidelity numerical methods for solving these partial 
differential equations. Furthermore, the equivalent relation between the electrostatic force density $f$ and the 
Maxwell stress tensor (MST) \cite{Gilson93} allows us to use the MST to compute the electrostatic force exerted only
on the surfaces of the membrane:
\begin{eqnarray*}
f & = & \bfM \cdot n  -k_B T \sum_{i=1}^N c_i ( e^{-q_i \phi/k_BT} -1), \\ 
\quad \bfM & = & -\frac{1}{2} \epsilon_s |\nabla \phi_s|^2 + \epsilon_s \nabla \phi_s \otimes \nabla \phi_s.
\end{eqnarray*}
It is worth noting that the charged membrane will subject to a self-force $f_0$ in the absence of the proteins. 
This self-force must be subtracted and thus the net traction applied on the membrane is $f-f_0$.

To ensure the accuracy of the numerical solution of electrostatic potential induced by singular point charges,
singular surface charge distribution, and mobile ions, we introduce a stable decomposition of the PB 
equation \cite{ZhouY2010a}, with which one can analytically solve the potential component induced by the singular 
charges. 
A molecular surface conforming finite element method is then used to
solve the regular component of the potential with which the interface conditions due to the surface charge 
distribution can be enforced exactly.  In order to avoid regenerating interface-conforming finite element meshes 
due to the membrane deformation in the iterations,
a Piola 
transformation \cite{MathematicalelasticityCiarlet} is adopted so that the elastic equation can be solved on the 
same reference configuration with varying surface traction. 

\begin{figure}[!ht]
\begin{center}
\includegraphics[height=3cm]{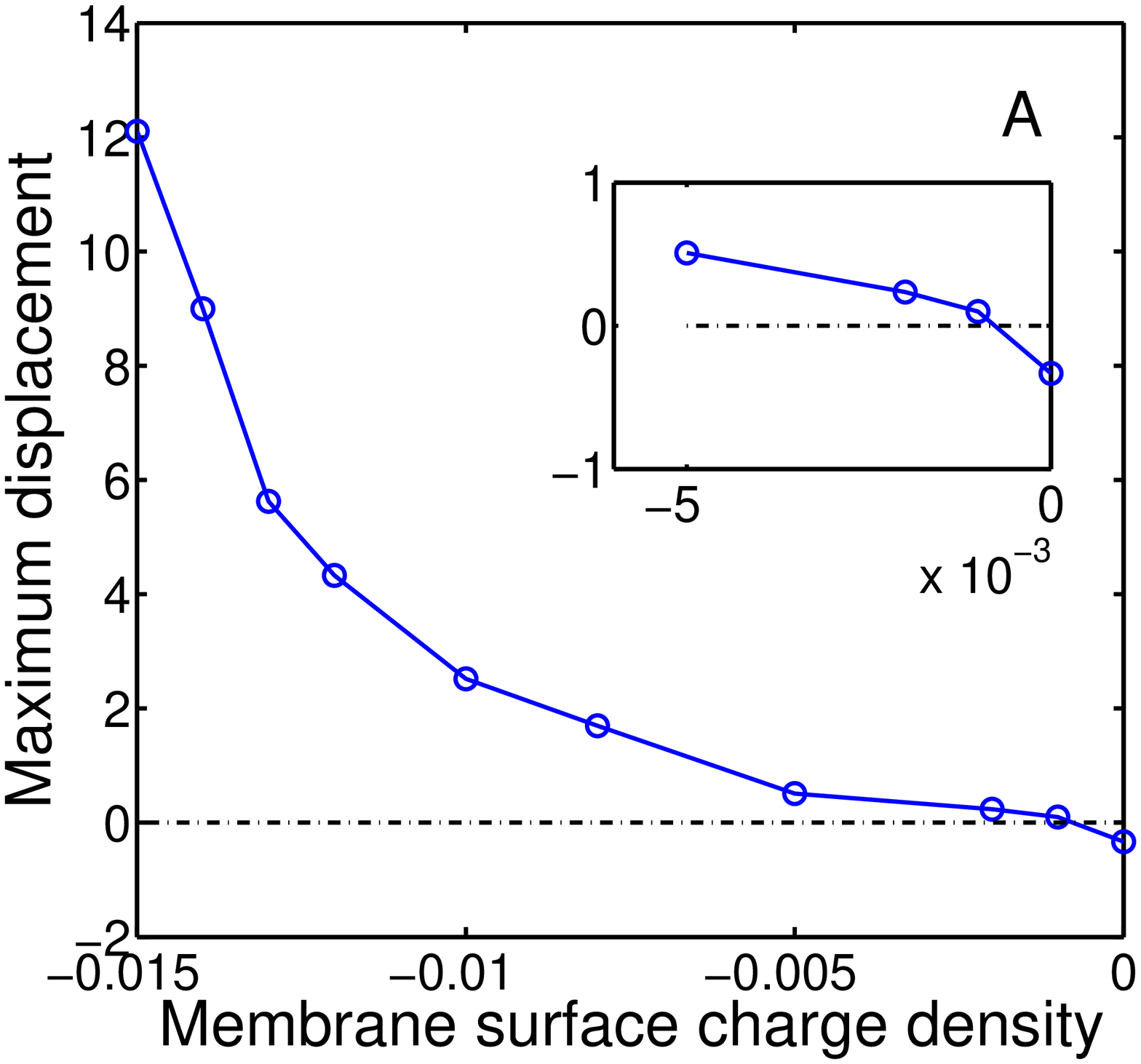}
\includegraphics[height=3cm]{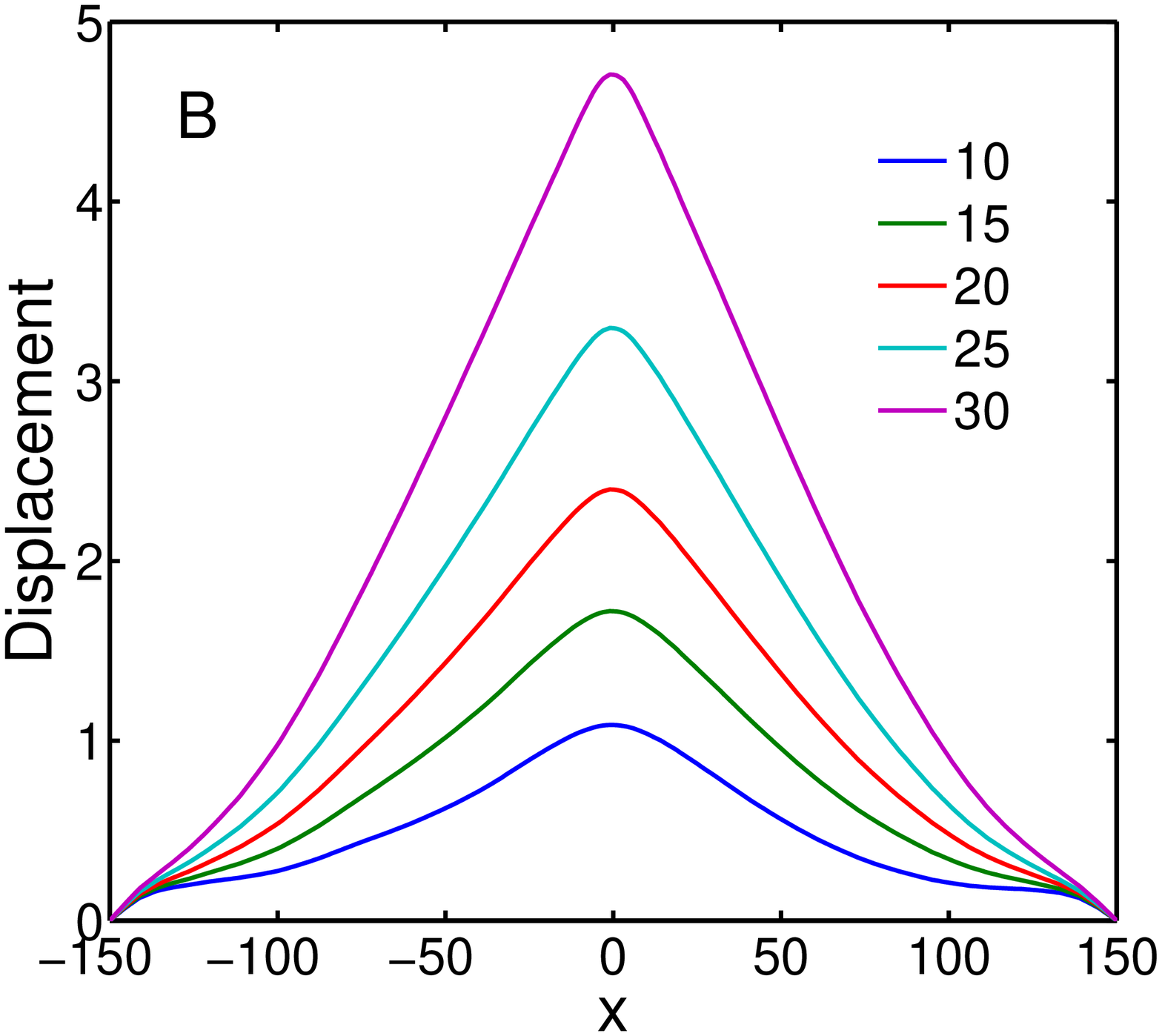} \\
\includegraphics[height=3cm]{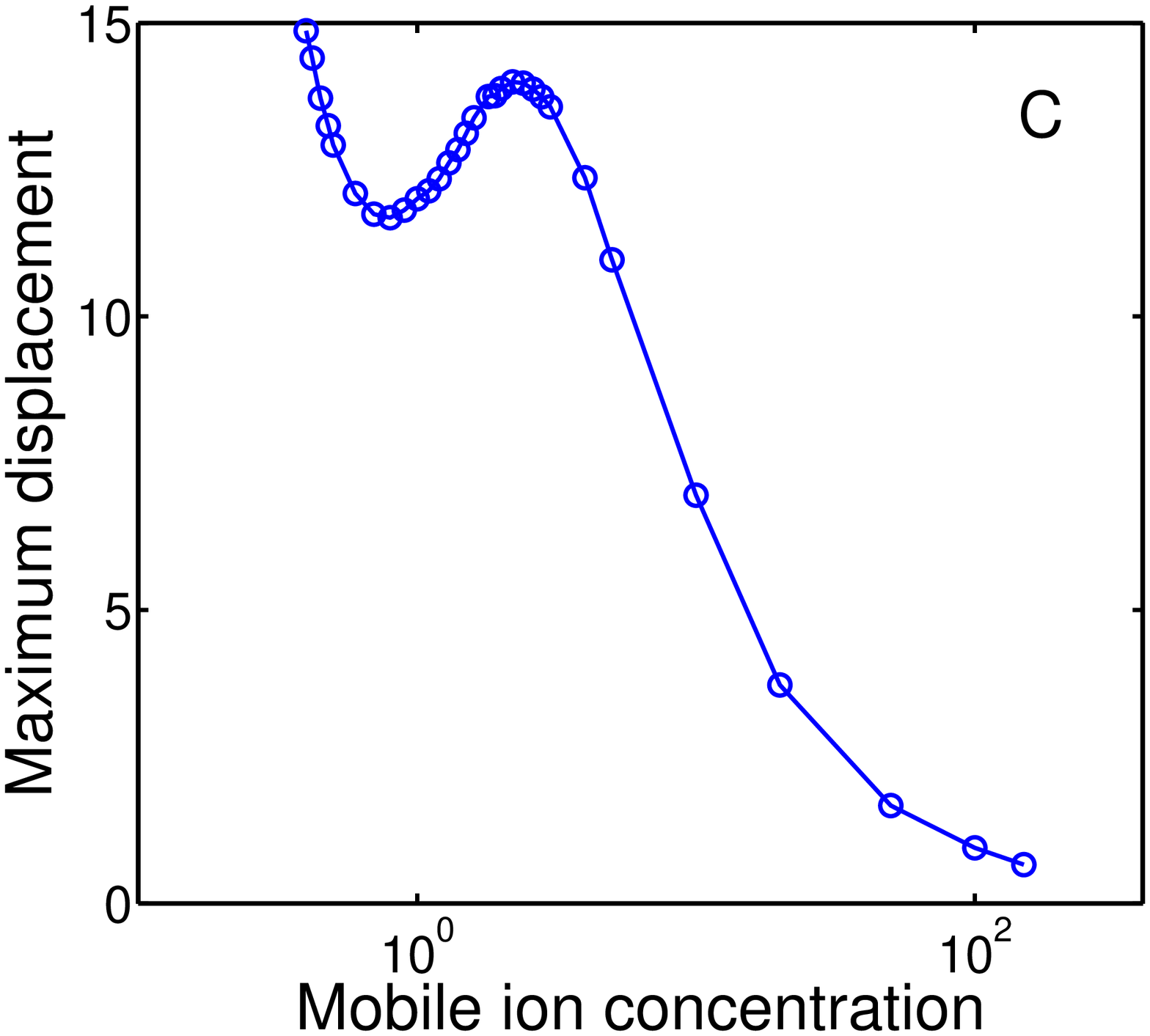}
\includegraphics[height=3cm]{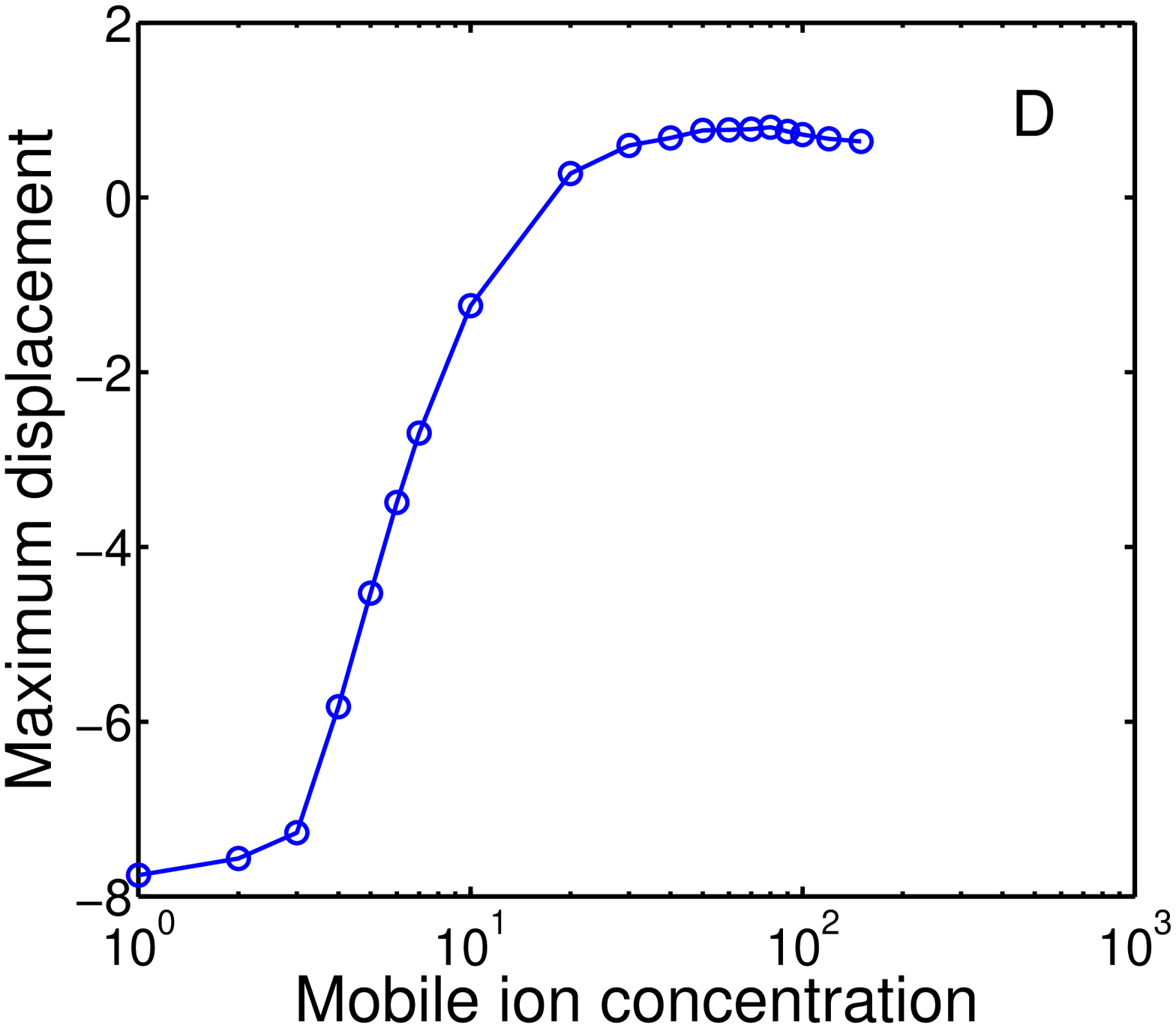}
\end{center}
\caption{Change of membrane displacement with different parameters for the test systems.
Unless specified particularly the charge of atom is $5e$, the mobile ion concentration is $150mM$, and the surface 
charge density assumes the default value. In the first system (A,B,C) the protein consists of a single atom 
centered at ($0,0,50$). In the second problem (D) the overall electrically neutral protein consists of 8 model 
atoms centered at ($\pm 6, \pm 6, 50$) and ($\pm 6, \pm 6, 56$). Every atom on $z=50$ and $z=56$ 
carries a charge of $2$ and $-2$, respectively. The center of the 
membrane is initially placed at ($0,0,0$) in both problems with its axle aligned with the $z$-axis.} 
\label{fig:model_problem}
\end{figure}

As a proof of the model and the numerical method, we present a test system in which a cylindrical membrane is buckled
by a model protein whose structure is adjustable. The membrane has a radius of 150\AA~and a thickness of 30\AA. 
The radii of atoms in the model protein are uniformly 10\AA. The default surface charge density $-0.0044e/$\AA$^2$
is computed by assuming that the membrane has the same composition of 30\% dioleoylphosphatidylserine (DOPS) 
and 70\% dioleoylphosphatidylcholine (DOPC) as studied in \cite{BloodP2006}\footnote{A replicated $ 5 \times 2$ system
has a surface area $251 \times 98$\AA$^2$, each unit having 36 lipids in its monolayer. Only 30\% of lipids have a 
unit negatively charged head group. The surface charge density is then 
$5 \times 2 \times 36 \times 0.3/(251 \times 98) = 0.0044$.}.
We fix the Young's modulus and the Poisson ratio of the membrane to be
$1pN/$\AA$^2$ and $0.3$, respectively \cite{TangY2006a}. 
The number, positions and charges of the atoms 
in the protein, the membrane surface charge density, and the mobile ion concentrations will be varied to validate the proposed 
model for simulating the membrane deformation under a wide range of electrostatic forces. Fig.~\ref{fig:model_problem} 
plots the membrane displacement as a function of stated parameters. 
The plots show that rich dynamics can take place in this electromechanical system and different types of electrostatic force 
compete. A large negative surface charge density results in a strong attractive interaction with the positively charged model protein, 
and gives rise to a positive displacement. In the absence of membrane surface charge, a small but negative 
displacement (Fig.~\ref{fig:model_problem}A) suggests that 
the total electrostatic force induced by the discontinuous dielectric permittivity and the ionic osmotic pressure 
is repulsive on the low dielectric material
in agreement with the analysis in \cite{ZhouY2007c,LiB2009a}. The attractive interaction gets stronger with the increase of the net 
positive charge of the protein, resulting in a monotonical increase of the membrane displacement (Fig.~\ref{fig:model_problem}B). 
The mobile ion concentration has a more complicated consequence in regulating the protein-membrane electrostatic interaction
(Fig.~\ref{fig:model_problem}C). The present of mobile ion at low 
concentration reduces the electrostatic potential and in turn the electrostatic force on the membrane surfaces, resulting a 
decease of the displacement. Nevertheless, the further increase of the mobile ion concentration from $0.8mM$ to $2.2mM$ 
leads to a slight increase of the membrane displacement. This suggests that over this small range of mobile ion concentration 
the decrease of the attractive force is dominated by the decrease of repulsive force components, resulting in an increase of 
attractive net force. This dominance reverses with the further improvement of the ion concentration, as seen in the decrease
of the displacement for all ion concentrations larger than $2.2mM$. In the second test system, the lower positively charged 
domain of the protein has an attractive interaction stronger than the repulsive interaction between the negatively charged 
upper domain and the membrane. One the other hand, the repulsive interaction
due to the discontinuity of the dielectric permittivity is also prominent under the weak screening of the mobile ions,
and thus a negative displacement as large as $-8$\AA~is seen in Fig.~\ref{fig:model_problem}D. Noticing that the dielectric
pressure (2nd term in Eq.(\ref{eqn:eforce_def})) is proportional to the square of the magnitude of the electric field while 
the $qE$ force (1st term in Eq.(\ref{eqn:eforce_def})) is proportional to $E$, this pressure reduces more quickly as the electric 
field weakens with the increase of ionic strength. At a sufficiently high ion concentration close to the physiological 
condition, a small positive displacement is found.  

\begin{figure}[!ht]
\begin{center}
\includegraphics[height=3cm]{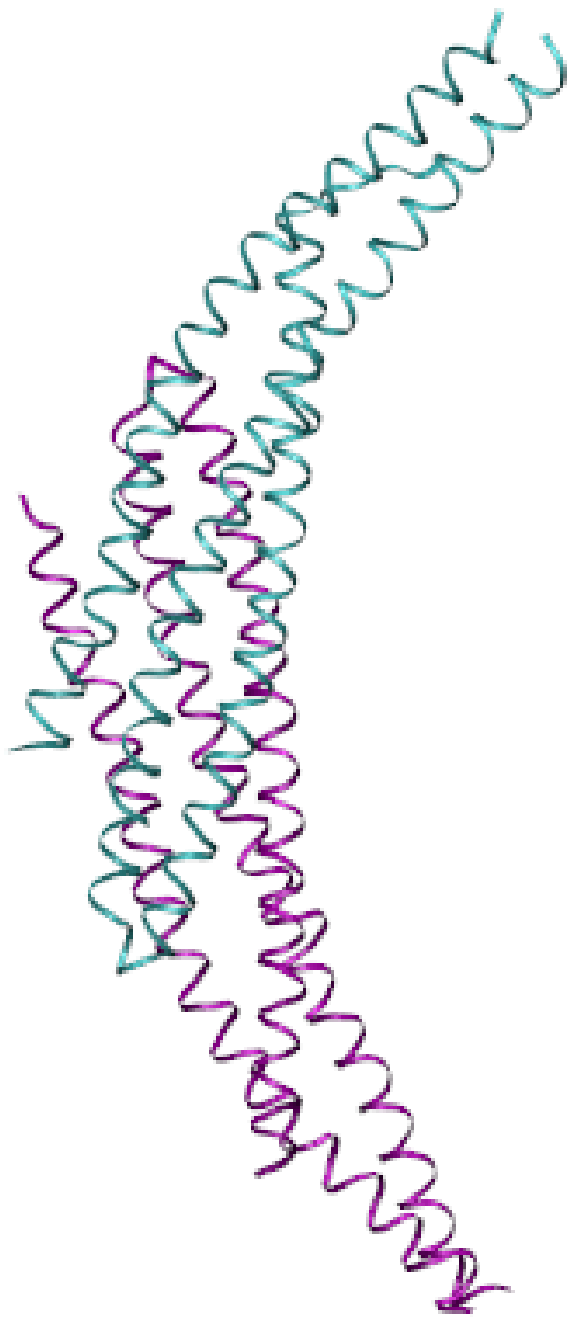} \hspace{5mm}
\includegraphics[height=3cm]{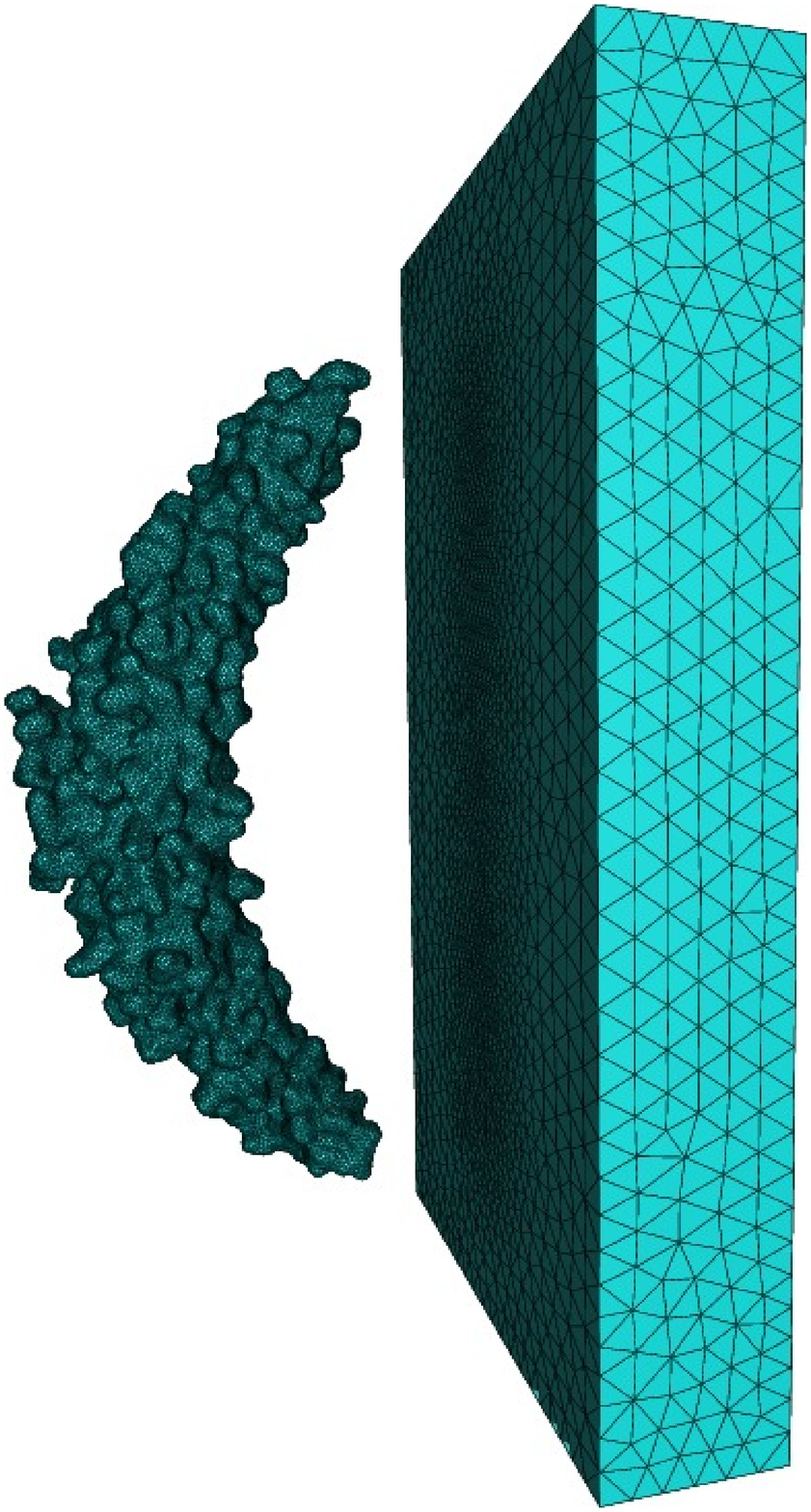}  \hspace{5mm} \includegraphics[height=3.2cm]{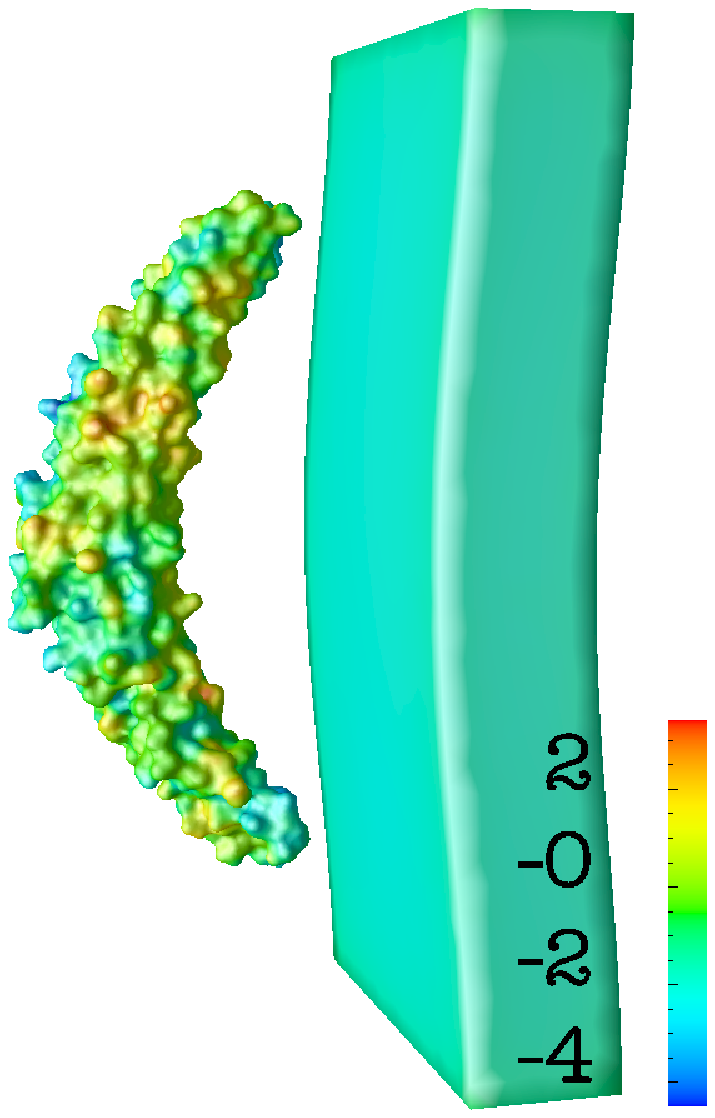} \\
\includegraphics[height=3cm]{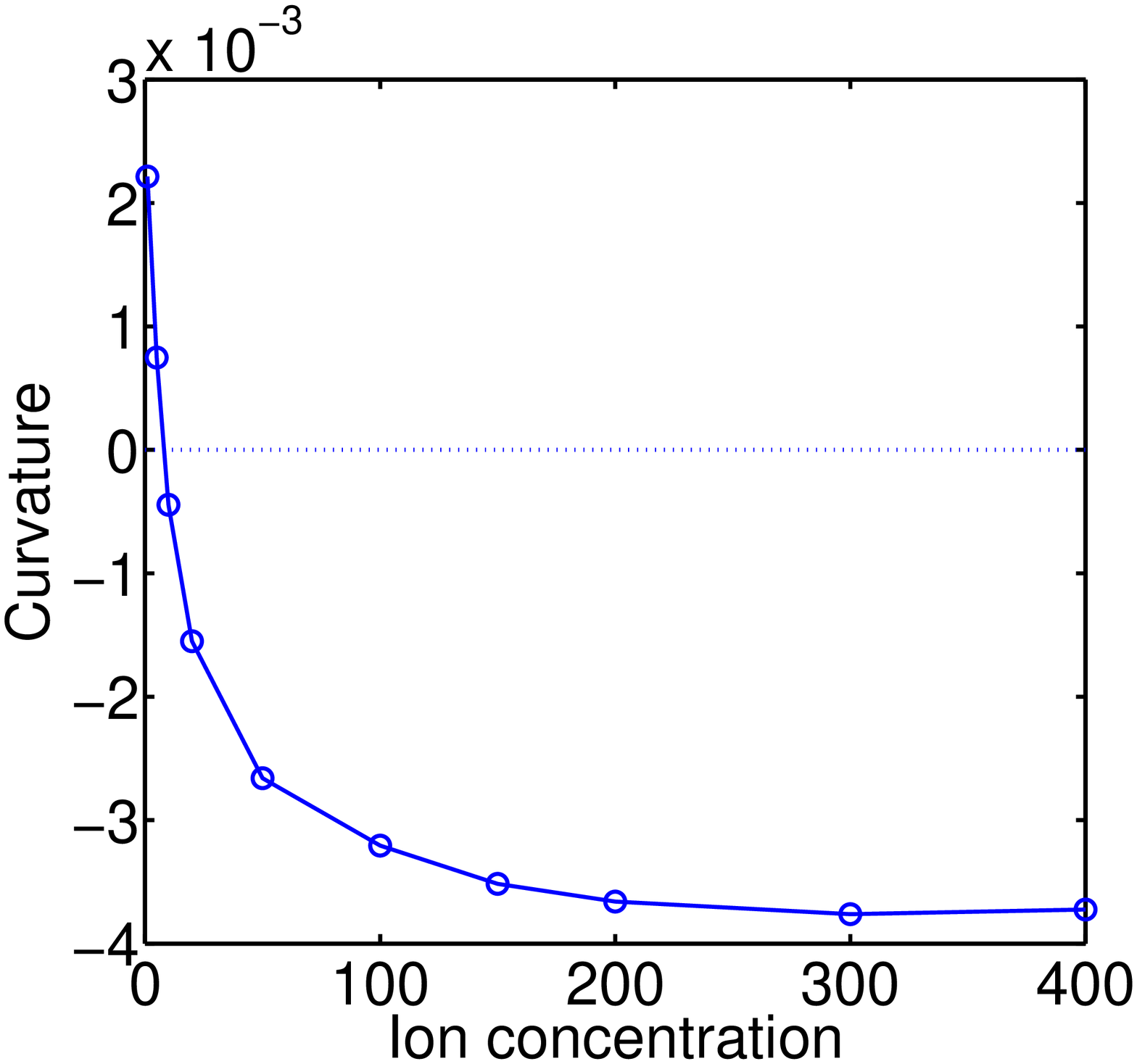} \hspace{4mm}
\includegraphics[height=3cm]{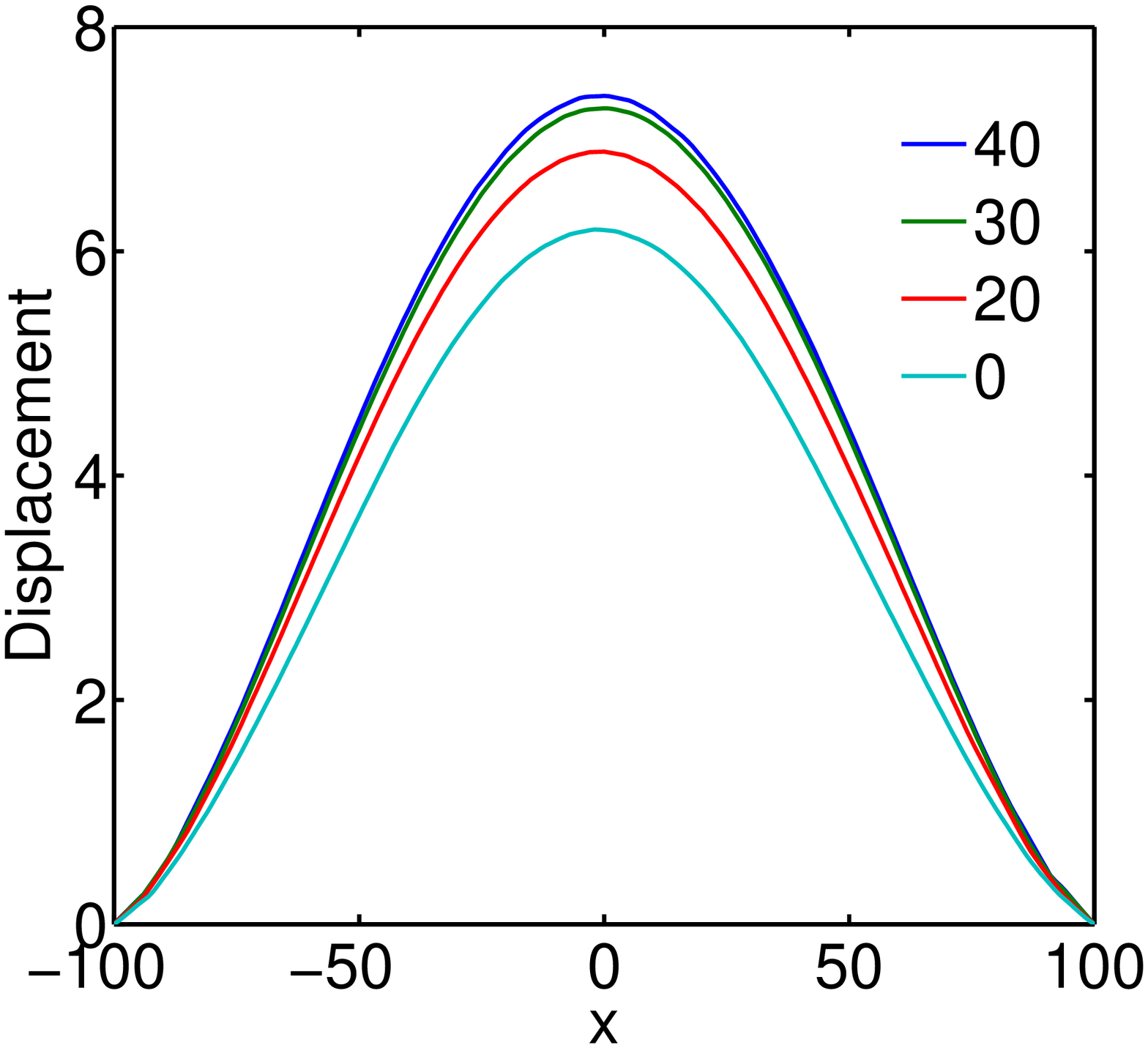} 
\end{center}
\caption{First row: The structure of protein 1I4D (Left); Surface triangulation of the BAR-domain protein-membrane complex (Middle) and 
the electrostatic potential on the protein and membrane surfaces (Right). Second row: Membrane curvature at various ion concentrations 
(Left) and membrane displacement at various initial separations $D$ (Right).}
\label{fig:BAR}
\end{figure}

We now investigate the membrane curvature induced by the BAR-domain protein (PDB ID: 1I4D) dimer and compare our results to the 
molecular dynamics simulations by Blood {\it et. al.} \cite{BloodP2006}. The membrane at free state is a rectangular cuboid 
whose size is $(x \times y \times z)=[-100,100] \times [-50,50] \times [-15,15]$\AA$^3$. The concave surface of the dimer has 
a radius of curvature $80$\AA. The coordinates of the dimer structure are rotated and translated such that the three 
atoms (GLN225:O,LEU149:HG,SER164:HG) are on the $x-z$ plane and the $z$-coordinates of the last two atoms are $0$. 
The $z$-distance between these two atoms and the top surface of the membrane, which is initially placed below the
dimer, is denoted by $D$. The top and bottom membrane faces are charged at the default value, and the other four 
faces are charge free. The two end faces in $x$-direction are fixed and thus homogeneous Dirichlet boundary conditions 
will apply. 
The periodic boundary condition is enforced in $y$-direction to reduce the numerical artifacts due to finite truncation. 
The results for the membrane displacement
and curvature, measured along the intersection between the top surface and the plane $y=0$, are shown in Fig.~\ref{fig:BAR}.
The dimer has a positively charged concave face and a net charge of $-10e$. This negative net charge gives rise to a strong 
repulsive $qE$ force to the negatively charged membrane surfaces when the ion screening is weak, which along with the repulsive 
dielectric pressure leads to a positive bending curvature of $2.2\times10^{-3}$/\AA. A transitional ion concentration is found 
at $8mM$, below which the bending curvature is positive while above which the curvature is negative. Similar to the second model
problem we tested above, this transitional ion concentration corresponds to the state in which the screened attractive interaction 
between the positively charged side of the dimer and the negatively charged membrane surfaces is balanced by the repulsive interaction 
between the negative charged side of the dimer and the membrane surfaces plus the dielectric pressure as well as the ionic pressure. 
The maximum curvature, $3.7 \times 10^{-3}$/\AA, occurs at an ion concentration of $200mM$, while the curvature at the physiological 
concentration, $100mM$ in this study, is $3.2 \times 10^{-3}$/\AA. These demonstrate that the present model is able to capture
the critical role of the mobile ions in mediating the protein-membrane interactions. Nevertheless, all these bending 
curvatures are smaller than the intrinsic curvature of the dimer ($1.25 \times 10^{-2}$/\AA). We speculate that a larger
curvature could be found at suitable initial separation between the membrane and the dimer. However, the simulations 
with various separations (c.f. Fig.~\ref{fig:BAR}) illustrate that the bending curvature becomes slightly smaller 
as the separation gets smaller. These observations suggest that the attractive electrostatic force is not sufficiently 
strong to generate a bending curvature as large as the intrinsic curvature of the dimer when the membrane is constrained at 
two $x$-ends. On the other hand, a charged membrane free of these constraints will move towards the dimer under the attraction 
and eventually contact with the two ends of the dimer; these two ends will then serve as the constraints for the bending 
of the membrane. This procedure has been observed in molecular dynamical simulations \cite{BloodP2006,ArkhipovA2008a}, 
where the two ends of the dimer will insert into the membrane after contacting with the membrane and further facilitate 
the bending of the membrane.

The results clearly demonstrate that our continuum electromechanical model efficiently captures the balance between the 
electrostatic energy and the elastic strain energy in the protein-membrane interaction. The model quantitatively
characterizes the dependence of the interaction on the solvation, partial charge distribution in protein, membrane 
composition and its surface charge distribution, and the membrane dielectric constant, some of which has been observed in previous
studies \cite{BloodP2006,ArkhipovA2008a}, but never described in a continuum framework. The present formalism is mainly limited by the 
fixed constraints on some un-charged membrane surfaces, which could be resolved by introducing the protein surface as a 
limit of membrane deformation \cite{ContactProblemsKO}. Future efforts to improve these approximations are critical to 
accurately describe the protein-membrane interactions for more complicated geometries. 
The formulation and examples presented here are expected to spur the interests of other investigators in molecular biology, 
mechanics, and mathematics. 

The authors thank Si Hang for the help with mesh generation. The research of YCZ is supported by Colorado State University. 
BZL is supported by the Chinese Academy of Sciences, the State Key Laboratory of Scientific/Engineering Computing, 
and NSFC(NSFC10971218). AAG acknowledges the support from the University of Texas Medical School at Houston.

\end{document}